\documentclass[aps,twocolumn,prl,showpacs,floatfix]{revtex4}

\usepackage{amsmath}
\newcommand{\texp}{\tau_\mathrm{exp}}
\newcommand{\texpO}{\tau_{\mathrm{exp},O}}
\newcommand{\tint}{\tau_\mathrm{int}}
\newcommand{\tintO}{\tau_{\mathrm{int},O}}

\usepackage{graphicx}

\begin{document}

\title{On the critical behavior of the specific heat in glass-formers}
\author{L.~A.~Fern\'andez} \affiliation{Departamento de F\'\i{}sica
Te\'orica I, Universidad Complutense, Av. Complutense, 28040 Madrid,
Spain.}  \affiliation{Instituto de Biocomputaci\'on y F\'{\i}sica de
Sistemas Complejos (BIFI). 50009 Zaragoza, Spain.}
\author{V.~Mart\'\i{}n-Mayor} \affiliation{Departamento de F\'\i{}sica
Te\'orica I, Universidad Complutense, Av. Complutense, 28040 Madrid,
Spain.}  \affiliation{Instituto de Biocomputaci\'on y F\'{\i}sica de
Sistemas Complejos (BIFI). 50009 Zaragoza, Spain.}
\author{P.~Verrocchio\footnote{Present address: Dipartimento di
Fisica, Universit\`a di Trento, via Sommarive 14, 38050 Povo, Trento,
Italy.}}  \affiliation{Departamento de F\'\i{}sica Te\'orica I,
Universidad Complutense, Av. Complutense, 28040 Madrid, Spain.}
\affiliation{Instituto de Biocomputaci\'on y F\'{\i}sica de Sistemas
Complejos (BIFI). 50009 Zaragoza, Spain.}  \date{\today}

\begin{abstract}
We show numeric evidence that, at low enough temperatures, the potential
energy density of a glass-forming liquid fluctuates over length
scales much larger than the interaction range. We focus on the
behavior of translationally invariant quantities.  The growing
correlation length is unveiled by studying the Finite Size effects. In
the thermodynamic limit, the specific heat and the relaxation time
diverge as a power law. Both features point towards the existence of a
critical point in the metastable supercooled liquid phase.
\end{abstract}
\maketitle 

The glass transition differs from standard phase transitions in that
the equilibration time of glass-formers (polymers, supercooled
liquids, colloids, granulars, super\-conductors, ...) diverges without
dramatic changes in their structural properties. Indeed elastic
neutron scattering experiments show that the Static Structure Factor
of polymers and supercooled liquids undergoes marginal modifications
when approaching the glass transition. Yet, their equilibration time
$\tau$, as obtained for instance from the $\alpha$ peak in the a.c.
dielectric susceptibility, grows by many order of magnitude close to
the transition\cite{DeBenedetti97}.  Reconciliating the two faces of
the glass transition is a major challenge for condensed matter
physics.

A general mechanism producing a divergence of the equilibration time
of an homogeneous system at finite temperature is the divergence of a
correlation length ({\em critical slowing
down}\cite{ZINNJUSTIN}). Slowness is due to the fact that
configurational changes need to propagate over increasingly large
regions (the critical origin of the Mode Coupling
singularity\cite{goetze:1992} has been recently
recognized\cite{biroli:2004}). Within this framework, a key problem is
identifying the quantity with the largest spatial fluctuations close
to the glass transition. Since equal-time correlation functions do not
reveal statistical fluctuations over large length scales, it has been
suggested that two-times correlation functions must be
studied\cite{biroli:2004,tarjus96,donati02,cugliandolo:2003,whitelam:2004}.
The goal is to extend beyond mean-field level the Mode Coupling view
of the glass transition as a purely dynamic phenomenon.

Strong space fluctuations of the relaxation properties ({\em dynamic
heterogeneities}) have been found
experimentally\cite{ISRAELOFF00,EDIGER00} and
numerically\cite{BERTHIER04}. However, the size of the correlated
domains is not larger than a few nanometers (a few angstroms, in
simulations).

Here we report numeric simulations for a fragile glass-forming
liquid\cite{Bernu87,Grigera01} showing large scale fluctuations in the
specific heat, an equal-time correlation function.  Thus, we claim
that the dynamical features of the glass-transition are due to
critical slowing down arising from a continuous phase transition
suffered by the metastable liquid. The difficulty in recognizing it is
due to the fact that standard scattering experiments are not devised
to detect spatial fluctuations in the energy density. Yet, a large
correlation-length can be studied through Finite-Size
effects\cite{FSSBOOK}.  Note that experiments in films\cite{FSS-FILMS}
and nanopores\cite{FSS-PORES} show that the glass transition changes
in samples with one or more dimensions of nanometric scale, although
it is difficult to disentangle Finite-Size Scaling from the effects of
the interaction with the substrate. However, the specific-heat of
toluene confined on pores of diameter 8.7\,nm \cite{FSS-PORES}, close
to its glass temperature, is significantly smaller than for bulk
toluene, which could signal a correlation length well above the
nanometric scale.

We study a 50\% mixture of particles interacting through the pair
potential
$V_{\alpha\beta}(r)=\epsilon[(\sigma_\alpha+\sigma_\beta)/r]^{12} +
C_{\alpha\beta}$, where $\alpha,\beta=A,B$, with a cutoff at
$r_\mathrm{c}=\sqrt{3}\sigma_0$. The choice $\sigma_B=1.2\sigma_A$
hampers crystallization. We impose
$(2\sigma_A)^3+2(\sigma_A+\sigma_B)^3+(2\sigma_B)^3=4\sigma_0^3$ where
$\sigma_0$ is the unit length.  Constants $C_{\alpha \beta}$ are
chosen to ensure continuity at $r_\mathrm{c}$. The simulations are at
constant volume, with particle density fixed to $\sigma_0^{-3}\,$ and
temperatures in the range $[0.897 T_\mathrm{mc}, 10.792
T_\mathrm{mc}]$, where $T_\mathrm{mc}$ is the Mode Coupling
temperature\cite{goetze:1992}.  We use periodic boundary conditions on
a box of size $L$ (which discretizes momenta in units of
$q_\mathrm{min}=2\pi/L\,$) in systems with $512,1024,2048$ and $4096$
particles.  From the Molecular Dynamics van Hove self-correlation
function\cite{Cavagna:2002}, one has for argon parameters,
$\sigma_0=3.4$\AA, $\epsilon/k_B=120$K and $T_\mathrm{mc}=26.4$K.  We
shall obtain equilibrium data below $T_\mathrm{mc}$.

We modify the Grigera-Parisi swap algorithm\cite{Grigera01} to make it
{\em local}, in order to keep the algorithm in the dynamic
Universality Class\cite{ZINNJUSTIN} of standard Monte Carlo (MC).  The
elementary MC step is either (with probability $p$) a single-particle
displacement attempt or (with probability $1-p$) an attempt to {\em
swap} particles. The swap is made by picking a particle at random and
trying to interchange its position with that of a particle of opposite
type, chosen at random {\em among those at distance smaller than $0.6
r_\mathrm{c}$}.  Detailed balance requires that the Metropolis test
include not only the energy variation but the change in the number of
neighbors. The swap acceptance is independent of system size and
grows from 0.74\% at 0.9$T_\mathrm{mc}$ up to 6\% at
$2T_\mathrm{mc}$. In this work we use $p=0.5$ (named {\em local swap}
from here on) and $p=1.0$ (named standard MC). The time unit $t_0$ is
$N/p$ elementary steps.

Although time correlators differ for different dynamics, when studying
static quantities (e.g. specific heat) the choice of Monte Carlo
dynamics is a matter of practical convenience. We study local swap
{\em time} correlators only in order to explore {\em static}
properties. Yet, the asymptotic decay of time correlators of two
dynamics belonging to a single dynamic Universality Class is given by
the same function (up to a rescaling) of the correlation
length\cite{ZINNJUSTIN}. Since standard and swap MC are both local and
share the same conservation laws, we expect the ratio of their
autocorrelation times, Eq.(\ref{TAUDEF}), to be essentially constant,
even if their time correlators differ at intermediate times.

We assume that equilibrium behavior can be meaningfully studied in a
{\em metastable} phase\cite{SPINODAL}(i.e. the equilibration time for
the metastable liquid phase is much smaller than the crystallization
time). At the lowest temperatures we simulate up to 400 independent MC
runs. In the analysis we only consider histories longer than 100
exponential autocorrelation times (see below). We use the Jack-Knife
method to estimate errors\cite{FSSBOOK}.

The study of a stochastic classical dynamics is mathematically
equivalent to the diagonalization of a quantum
Hamiltonian\cite{PARISIBOOK}, related to the Fokker-Planck (FP)
operator, whose eigenvalues yield the relaxation time
scales. Translational invariance of the FP operator induces a
classification of physical quantities according to their wave vector,
$Q$.  The small $Q$ density fluctuations are ruled by the hydrodynamic
law stemming from mass conservation $\tau(Q) \propto
Q^{-2}$\cite{hansen}, which become singular at $Q=0$. Yet, the
dynamics of the translationally invariant quantities ($Q=0$ sector)
may be studied for non conserved quantities such as the potential
energy.

A generic observable $O$ is said to belong to the $Q$-sector if it
transform as $O \rightarrow \mathrm{e}^{\mathrm{i} \vec Q\cdot\vec
\Delta} O \,$ under an uniform displacement of the coordinates, $\vec
r_i =\vec r_i +\vec \Delta$.  Well known example are the Fourier
transforms ($V_{kj}\equiv V({\vec r_k -\vec r_j})$): 
\begin{equation}
\rho(\vec Q)= \frac{\sigma_0^{-3}}{N}\sum_{j=1}^N \mathrm{e}^{\mathrm{i} \vec
r_j\cdot \vec Q},\,\rho_e(\vec Q)= \frac{\sigma_0^{-3}}{2N}
\sum_{j,k\neq j}
\mathrm{e}^{\mathrm{i} \vec r_j\cdot \vec Q} V_{kj}.
\end{equation}
Note that $\rho(0)$ is the (conserved) particle density, while
$\rho_e(0)$ is the (non-conserved) potential energy density, which we
call $e$ (the internal energy is then $\frac32 k_\mathrm{B} T +\langle
e\rangle$).

Multiplying densities with wave-vectors $\vec q$ and $-\vec q$ yields
translationally invariant observables:
\begin{equation}
{\cal S}(\vec q) = \rho(\vec q) \rho(-\vec q)\ ,\quad {\cal S}_e(\vec q)=
\rho_e(\vec q) \rho_e(-\vec q)\,.\label{SQDEF}
\end{equation}
The mean value of ${\cal S}(\vec q)$ is the static structure factor
S$(\vec q)$, while $\langle {\cal S}_e(0)\rangle$ is related to the
constant-volume specific heat, $C_V$ as $T^2C_V= N
(\sigma_0)^6\left[\langle {\cal S}_e(0)\rangle - \langle e\rangle
^2\right]\,.$ For every observable, $O$, we consider the normalized
time correlator
\begin{equation}
%C_O(t)=\frac{\langle O (0) O(t)\rangle-\langle O\rangle^2}{\langle
%O^2\rangle-\langle O\rangle^2}\,.\label{CODEF}
C_O(t)=(\langle O (0) O(t)\rangle-\langle O\rangle^2)/(\langle
O^2\rangle-\langle O\rangle^2)\,.\label{CODEF}
\end{equation}
The Theory of Critical Phenomena\cite{ZINNJUSTIN} suggest that, at
{\em very long} times, time correlators decay
exponentially\footnote{Stretched exponentials fit nicely intermediate
times\cite{DeBenedetti97}.}. We consider two autocorrelation
times\cite{SOKALLECTURES}, the integrated time $\tintO$ and the
exponential time $\texpO\,$:
\begin{equation}
\tintO=\int_0^{\infty} C_O(t)\,\mathrm{d}\,t\ ,\quad
C_O(t)\underset{t\to\infty}\longrightarrow
\mathrm{e}^{-t/\texpO}\,,\label{TAUDEF}
\end{equation}
with different meanings. The $\tintO$ of $Q=0$ quantities is related
to transport properties (e.g. viscosity is $\tint$ for some components
of the energy-stress tensor\cite{hansen}). The $\texpO$ is the longest
characteristic time and depends only on the $Q$-sector to which $O$
belongs.

\begin{figure}
\includegraphics[angle=270,width=1.0\columnwidth]{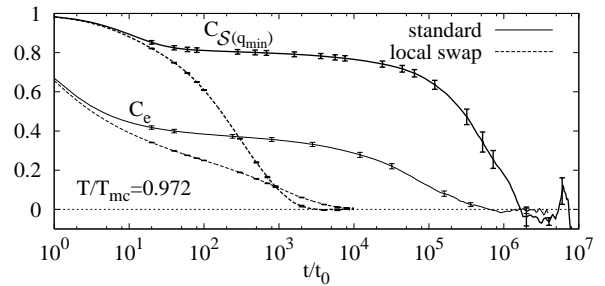}
\caption{Correlators for two translationally invariant quantities: the
potential energy density, $e$, and the square of the density
fluctuation, ${\cal S}(q_\mathrm{min})$, at the minimal momentum allowed by
the boundary conditions ($1024$ particles, below $T_\mathrm{mc}$). The
local swap algorithm decorrelates faster than standard
MC.\protect{\label{FIG1}}}
\end{figure}

\begin{figure}
\includegraphics[angle=270,width=1.0\columnwidth]{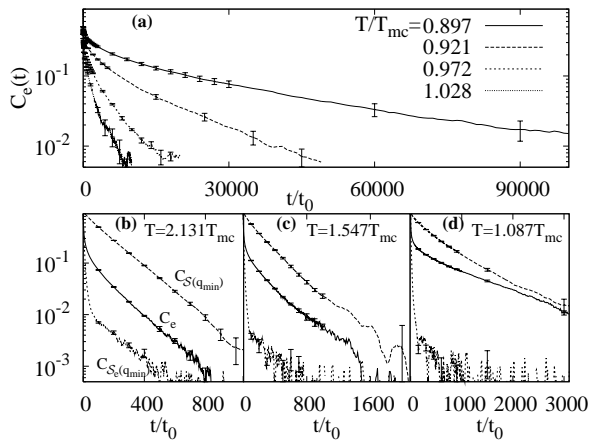}
\caption{Correlators for $N=1024$ particles ($N=2048$ at $T/T_{\mathrm
mc}=0.897$) with the swap algorithm.  Panel (a): temperature variation
of the correlator of $e$, close to $T_\mathrm{mc}$.  Panels (b)-(d):
correlators of several $Q=0$ quantities. Asymptotically, three
parallel straight lines should be found, with slope $-1/\texp$. At the
highest temperature, (b), this is clearly observed, as well as in
(c). In (d) we do not have enough independent measurements to see
clearly this common slope.  \protect{\label{FIG2}}}
\end{figure}

In Fig.~\ref{FIG1} we show the time correlator for the potential
energy density and ${\cal S}(\vec q_\mathrm{min})$. While at short
times standard MC and local swap dynamics yield the same correlators,
the swap does not present the Mode Coupling plateau (hence
significance of $T_\mathrm{mc}$ for swap MC is unclear), allowing a
better study of the long time decay.  Following this decay is
difficult. The numerical (or experimental) effort to obtain the
correlator $C$ with prescribed accuracy grows as $C^{-2}$ when $C\to
0$. In fact, most of previous work was confined to $C>0.1$, while we
are able to explore the range $0.01<C<0.1\,$ (Fig.~\ref{FIG2}).

In Fig.~\ref{FIG2} we show a semilogarithmic plot of the time
correlators versus time. For long times, a straight line of slope
$-1/\texp$ should be found. In Fig.~\ref{FIG2}(a) we show that $\texp$
obtained from $e$ grows fast near the Mode Coupling temperature. The
lower panels of Fig.~\ref{FIG2} confirm that $\texp$ is actually a
property of the $Q=0$ sector.  When our statistics are good enough to
follow the correlator for three decades, the estimate for the
exponential time does not depend on the chosen correlator. At low
temperature one must be content with choosing the most convenient
observable to extract $\texp$. Every observable within a symmetry
class has some overlap with the slowest mode of that class. The one
which shows the smaller slope at shorter times is closer to the
slowest mode, allowing more precise studies of $\texp$.

\begin{figure}
\includegraphics[angle=270,width=1.0\columnwidth]{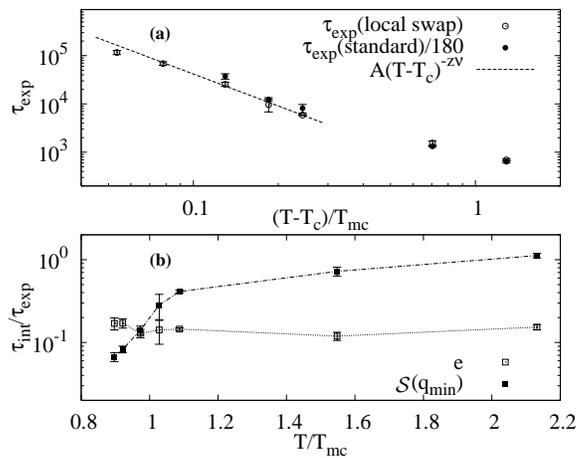}
\caption{Panel (a): $\texp$ vs. reduced temperature (line is a fit of
the swap data to a critical divergence).  Note Universality with
respect to changes in the dynamics (the two $\texp$ differ in a
constant factor). With standard MC one is restricted to $T\geq 0.973
T_\mathrm{mc}$.  (b): ratio $\tint/\texp$ for swap dynamics. The
participation of ${\cal S}(q_\mathrm{min})$ in the slowest mode
decreases upon approaching $T_\mathrm{c}$, while the share of $e$
slightly increases. We use $N=1024$ but for $T/T_\mathrm{mc}=0.897$,
where $N=2048$.  \protect{\label{FIG3}}}
\end{figure}

The growth of $\texp$ is shown in Fig.~\ref{FIG3}(a). Close to a
critical point, one expects a power law divergence of $\texp$. This
hypothesis accounts for our data, with an estimate for the critical
temperature $0.84 T_\mathrm{mc} < T_\mathrm{c} < 0.89
T_\mathrm{mc}\,.$

Universality requires that $\texp\propto A (T-T_\mathrm{c})^{-z\nu}$,
with $T_\mathrm{c}$ and the combination of critical exponents $z\nu$
independent of the dynamics.  Interestingly enough, in
Fig.~\ref{FIG3}(a) we find that the ratio of $\texp$ for the swap and
the standard MC dynamics is constant within errors, in spite of the
enormous difference on their short-time behaviors.  

A crucial issue is identifying the quantity with largest spatial
fluctuations. For local dynamics, this quantity is closely related to
the slowest mode (whose relaxation is purely exponential).  Closeness
can be quantified by the ratio of the integrated autocorrelation time
to the exponential one, Eq.(\ref{TAUDEF}). The challenge is to find
the observable that maximizes this ratio in the $Q=0$ sector
($\tint/\texp=1$ only for the slowest mode). In Fig.\ref{FIG3}(b) we
show the ratio $\tint/\texp$ for the potential energy and for ${\cal
S}(q_\mathrm{min})\,$, obtained with the swap dynamics. We observe
that ${\cal S}(q_\mathrm{min})$ seems a very good candidate at high
temperatures but the ratio $\tint/\texp$ sinks near $T_\mathrm{mc}\,$.
Fortunately, the potential energy displays a modest but constant ratio
(even slightly increasing at low temperatures). This is a strong
indication that critical behavior can be investigated in the
fluctuations of the potential energy, namely the specific heat, and
that a diverging correlation length would show up in four-particle
correlators.

\begin{figure}
\includegraphics[angle=270,width=1.0\columnwidth]{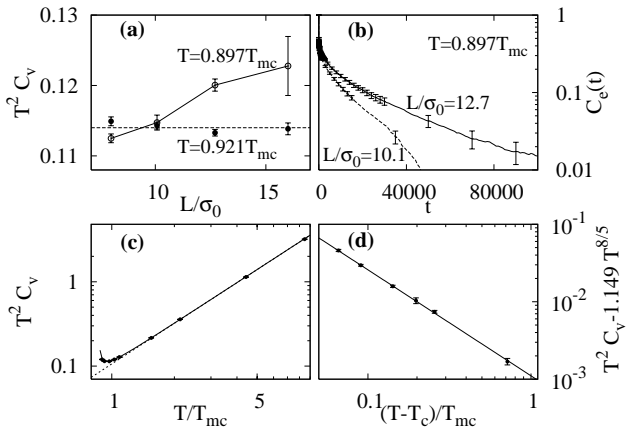}
\caption{Panel (a): unlike internal energy (not shown), the specific
heat is a growing function of the simulation box size, until the
system becomes much larger than the correlation length\cite{FSSBOOK}.
This is visible for $L<12 \sigma_0$ at $T=0.897T_\mathrm{mc}$, but not
at $T=0.921T_\mathrm{mc}\,$.  The correlation length grows quickly
close to $T_\mathrm{c}$, which is confirmed by the time correlator for
the potential energy (b).  Panel (c): infinite volume specific-heat
vs. temperature. The dashed line is the $T^{8/5}$ Rosenfeld-Tarazona
law. Deviations from the $T^{8/5}$ law can be fitted by a critical
divergence (full line in (d)) yielding
$T_\mathrm{c}=0.86T_\mathrm{mc}\,.$\protect{\label{FIG4}}}
\end{figure}

For the local dynamics studied here, diverging autocorrelation {\em
times} suggest diverging correlation {\em lengths}, that should show
up in static properties as well.  Since the order parameter is
unknown, a direct measure of the correlation length is difficult, but
one may detect it indirectly through Finite-Size
Scaling\cite{FSSBOOK}. The time correlators have taught us that energy
fluctuations are promising candidates. We studied, Fig.~\ref{FIG4}(a),
the specific heat dependence on the size of the simulation box,
$L$. Unfortunately, small systems crystallize quickly below $2.13
T_\mathrm{mc}$. On the other hand, the metastable liquid can be
studied with $N\!=\!512$ particles, using local swap, down to
$T\!=\!0.897 T_\mathrm{mc}\,$.  Up to $T\!=\!0.921 T_\mathrm{mc}$ no
finite-size effects are detected.  However, for $T\!=\!0.897
T_\mathrm{mc}$, a noticeable growth of the specific-heat with $L$ is
found up to $L=12\sigma_0$ ($\sim 4$nm for argon parameters). This
length is comparable with the experimental domain size for cooperative
dynamics\cite{EDIGER00,ISRAELOFF00}, and well above previous
simulations\cite{BERTHIER04}. The $\texp$ show a similar effect,
Fig.~\ref{FIG4}(b).

The next step is the study of critical behavior in the infinite-volume
specific heat itself, displayed in Fig.\ref{FIG4}(c). Generally
speaking, critical divergences for the specific heat are difficult to
study numerically due to the presence of a large non-critical
background (see e.g. Ref.\cite{BALLESTEROS98}).
Fortunately, in our case the background is given by the
Rosenfeld-Tarazona law\cite{ROSENFELDTARAZONA}, $T^2
C_\mathrm{V}\propto T^{8/5}\,,$ which should be followed by a
non-critical dense fluid at low temperatures. We have checked that
from $T=2 T_\mathrm{mc}$ to beyond $10 T_\mathrm{mc}$ the $T^{8/5}$
law is extremely accurate.  Interestingly, at lower temperatures
(where the agreement with the $T^{8/5}$ law should be still better in
absence of criticality) {\em deviations} start to be significant. They
are well described by a power law divergence $\propto
(T-T_\mathrm{c})^{-\alpha}\,$. To estimate errors in $T_\mathrm{c}$
and $\alpha$ is difficult.  Excluding the two extremal points in
Fig.\ref{FIG4} (c), the fit yields $\alpha=0.9$ and $T_\mathrm{c}=0.86
T_\mathrm{mc}$. Including any of the extremal points in the fit,
$\alpha$ slightly grows. Taking into account the bound $\alpha<1$
imposed by the continuity of the internal energy, we estimate
$0.9<\alpha<1.0$ and $0.84 T_\mathrm{mc} <T_\mathrm{c}< 0.86
T_\mathrm{mc}$.

In summary, we studied the equilibrium static and dynamic properties
of a model of a fragile supercooled liquid, with emphasis on the
dynamics of translationally invariant quantities. We claim that
critical slowing down is behind the structural arrest of glass
formers, which have then universal properties.  We find that time
correlators with a complicated structure relax exponentially at very
long times.  The study of the time correlators is a powerful,
unprejudiced method of identifying the physical quantities suffering
fluctuations over the largest length scale. The potential energy,
rather than density fluctuations, emerges as the candidate for the
study of this critical phenomenon. Experiments measure the constant
pressure specific-heat, that is obtained from the constant volume
one by adding a term that is smooth in the absence of critical density
fluctuations (not found here). The critical temperature obtained
from the divergence of the specific heat and of the autocorrelation
times lies in the range $T/T_\mathrm{mc}=0.83$---$0.88\,.$ The study
of the dynamics of translationally invariant quantities appears as a
challenge to experimentalists. While measurements of the frequency
dependence of the specific heat\cite{CARRUZZO04,NAGEL85} are an
appealing possibility to estimate the potential energy relaxation
time, the correlation-length could be studied by Finite-Size Scaling
of the specific-heat and of relaxation times in films\cite{FSS-FILMS}
or in larger pores than previously used to confine
glass-formers\cite{FSS-PORES}.

We thank R. de Nalda, T.S. Grigera, G. Parisi and C. Toninelli for
discussions.  P.V. was supported by the EC (contract MCFI-2002-01262).
We were partly supported by MEC (Spain), through contracts
BFM2003-08532, FIS2004-05073 and FPA2004-02602.  The total CPU time
devoted to the simulation (carried out at BIFI PC clusters) amounts
to 10 years of 3 GHz Pentium IV.

\bibliography{biblio}

\end{document}